New ionic liquids of ultralow proton activity.

C. Austen Angell,
School of Molecular Sciences, Arizona State University, Tempe, AZ 85202

The route to superacidic ionic liquids by proton transfer from molecular superacids to weak molecular bases like pentafluoropyridine, has been described in recent work. Here we consider the process of making superbasic ionic liquids by a similar procedure, and encounter the problem of finding molecular bases of high enough basicity to accept protons from weak acids to form ionic liquids of both high basicity and high ionicity. The consequence is that no ionic liquid with proton activity outside the aqueous base limit of pH =14, has yet been made. The problem is resolved by considering the six possible types of proton transfer processes, selecting the process of transfer from weak molecular acids to ultrabasic *anions* (partnered with alkali metal cations) to create new anions of intermediate basicity, and then replacing the alkali metal cations with reduction-resistant organic cations using metathesis. In many cases the alkali metal salts of target anions are commercially available, and only the metathesis reaction is needed to obtain the basic ionic liquid. With this approach we show how to extend the possible proton activity range down, by some 25 orders of magnitude, to reach $pK_a$ values of order 40, with even higher values possible in principle. ("There's plenty of room at the bottom" - with apologies to Richard Feynman). A thermodynamic route to the assessment of key basicity levels on the water scale, i.e. relative to the $H_3O^+/H_2O$ donor/acceptor level, is described, and is verified by application to the case of pure $H_2SO_4$.

## 1. Introduction

Protic ionic liquids have been given increasing attention in recent years, partly because of the simplicity of their synthesis and partly because of interest in the extra dimension of tunability they possess by virtue of the dependence of the proton activity on the choice of acid and base between which the proton transfer occurs(*1-7*). Of much assistance to the present author in this respect has been the manner of organizing the available information on Bronsted acids and bases proposed long ago by Ronald Gurney for the standard states of acids and bases in water(*8*). Gurney coupled protonated and unprotonated states together as "occupied" and "vacant" proton states of the same chemical entity, which might otherwise be referred to as that same entity in its proton-donating and proton-accepting roles, (analogous to reduced and oxidized states of a redox species in an electron free energy level diagram). Then different couples were arranged in an energy level diagram in which the separation of levels is quantified by the free energy of the proton transfer reaction as the proton "falls downhill" from the high chemical potential, occupied state on a donor in the solvent, to the lower free energy vacant state on some chosen acceptor - also in the same solvent (water).

There is of course a problem in applying Gurney's approach to ionic liquids, quantitatively since there is no unique or obvious single ionic liquid to serve as solvent for the donor/acceptor couples, nor are there data for the free energies of those couples in such a chosen solvent. However, there is room for a cruder approach which is very helpful from an organizational point of view. This is to make the assumption that the free energy of hydration of each of the cation and the anion in a given protic ionic liquid, as it passes from the anhydrous (pure PIL) state to the aqueous standard state, is the same. While this would obviously be wildly incorrect for a salt like lithium chloride, it is surely much less incorrect for a weak field ionic liquid like hydrazinium trifluoroacetate, or diethylmethylammonium methanesulfonate, both of them "good" protic ionic liquids (PILs). Indeed, this view is supported by the recent observation of the great difference in cooling behaviors between lithium chloride and hydrazinium trifluoroacetate aqueous solutions of about the same molar concentration(*9*). Where the former dismantled all structures responsible the anomalies of the supercooled water state, the hydrazinium trifluoroacetate left them intact, sufficiently to reveal the much-discussed liquid-liquid (HDL-LDL) transition for supercooled water (see Figure 1 of ref. (*9*))

Thus, with the admittedly disconcerting assumption that the free energy differences between occupied/vacant states of a given molecular acid and those of a given molecular base might be semi-quantitatively the same in anhydrous as in aqueous standard states, we have used the $pK_a$ data from aqueous solution studies, their theoretical extensions, and Hammett acidity function data, to construct proton free energy level diagrams (or Charts(8)) for ionic liquids. These have proven to be semiquantitatively successful in correlating experimentally observed thermodynamic properties such as excess boiling points(*1*) and redox potentials(*10, 11*) kinetic properties such as glass transition temperatures and viscosities(*3*), and spectroscopic properties such as $^{1}$H-N and $^{15}$N NMR chemical shifts(*6, 12-14*).

In this paper we will use this Gurney style diagram to summarize the range of behavior to be expected with systems for which aqueous $pK_a$ data are available, either by direct measurement or correlation, or by theoretical argument, and then confront the problem that arises when the conventional proton transfer (from molecular acid to molecular base) can no longer be used for synthesis due to lack of suitable molecular bases. This limit occurs roughly at the $pK_a$ value of 9 for a proton transfer across a large enough energy gap to still give a highly ionic product. The limitation is due to the lack of small molecular bases with $pK_a$ values less than 16 on the water scale to serve as the proton acceptors from weak acids. The so-called superbase, DBU, for instance, only has a $pK_a$ value of 14 on the water scale. Since a $\Delta pK_a$ of some 10 units or more is needed for high ionicity(*6, 15*), this means that the weakest acid suitable for proton donation would be acetic acid, pKa = 4.7, and then the midpoint $pK_a$ for the product [DBUH+][OAc-] would only be ~9, see Figure 1 below. This is not even outside the standard aqueous range of 0-14.

Since there is a natural interest in ionic liquids with basicities lower than this we will then consider how to extend the proton activity of an ionic liquid to much lower values and how to quantify the proton activities of the ionic liquids that might be generated.

In Chart 1 we reproduce the Gurney-inspired proton energy level diagram for the range from superacids as the donor species down to water itself. This is considerably updated from those published elsewhere (most recently 2012(*7*)), so as to reorganize the superacid/acid boundary and include recent information on the relative proton potentials of (transient) $HAlCl_4$ and $HSbF_6$ from $^{15}N$ and $^{1}H$ NMR and conductivity studies(*12, 13, 16*). The energies in eV assigned to the levels are relative to that for the $H_3O^+/H_2O$ level as reference and are calculated from the $pK_a$ values, using the simple relation $E(gap)/eV = \Delta pK_a/0.059$, explained later. They are numerically the same as the potentials, in volts, that drive the proton transfers.

The proton transfers of the type needed to produce ionic liquids of both superacidic character and high ionicity, are indicated by the top arrow in Chart 1, (Gurney-inspired). Even before the 2012 diagram was published(*7*), the superacidic character of transient $HAlCl_4$ and $HAl_2Cl_7$ in the presence of the very weak base pentafluoropyridine, had been established by the high ionic conductivity of the product $PFP^+Al_2Cl_7^-$(*17-19*). This is a stable liquid at ambient temperature, unlike the unstable toluenium tetrachloroaluminate PIL (reported in a classic 1955 paper by H.C. Brown and his coworker Pearsall(*20*)), which only exists below -45ºC. Above -45ºC, in the latter study, a disproportionation to release one mole each of HCl and PFP occurs, leaving a more stable PIL with the even weaker base anion $Al_2Cl_7^-$. More details on this interesting case, some analogs, and their practical advantages as benign (low vapor pressures) superacids, are given elsewhere (17). Other updates of the earlier charts are due to the incorporation of data in the superacidic domain of chart 1 from now-classical sources like the papers of Hammett(*21*) and the volume by Olah (*22*) supported and extended by recent works on ordering of acidities by $^{1}H$ and $^{15}N$ NMR studies(*12, 13, 16, 23*)

In this paper our main interest is in developing the lower proton activity part of the Gurney Chart which, at least in principle, extends down to proton activity levels more than 20 orders of magnitude below the limit of Chart 1(upper part, pKa < 14), more than doubling its range and reaching $pK_a$ values of about 40, vs 14 for the standard state value of the $H_2O/OH$- level in aqueous solutions (somewhat higher for saturated aqueous hydroxide or ionic liquid hydroxide couples). We commence with some details on what has limited the exploration of this high basicity range in previous ionic liquid studies. In essence it is the limitation imposed by the conventional Brønsted acid to Brønsted base proton transfer process by which, till now, all protic ionic liquids have been generated. The problem lies in the unavailability of small molecule proton acceptors of high basicity.

There are a number of imides and phosphazine molecules described in the literature as superbases(*24*), which have been valuable as reducing agents in organic chemistry but their $pK_a$ values, determined on the water scale, are generally not sufficient to generate an ionic liquid with an effective $pK_a$ value that is below that of the $H_2O/OH^-$ level of the Gurney type diagrams that have been published to date. Above we estimated that the $pK_a$ value of one of the most basic

**Chart 1**

**Extended Gurney style proton energy level diagram**

| Superacidic Electrolytes | Occupied proton level | Vacant proton level | $pK_a$ ($H_0$) ref $H_3O^+/H_2O$ | $\varepsilon^o$/eV vs $H_3O^+/H_2O$ | Refs and questions |
|---|---|---|---|---|---|
| hexafluoroantimonic | $HSbF_6$ | $SbF_6^-$ | -21.6 ($H_0$), -23.6 NMR | 1.3, 1.4 | a,b |
| Bis(triflide, HTFSI | $(CF_3SO_3)_2NH$ | $(CF_3SO_3)_2N$ | -18 ($H_0$), -19 NMR | 1.1, 1.1 | a.b |
| Perchloric acid | $HClO_4$ | $ClO_4^-$ | -16 ($H_0$), -16.1(NMR) | 0.95 | a,b |
| Triflic acid, HTf | $CF_3SO_3H$ | $CF_3SO_3^-$ | -14.6 ($H_0$), 14.6 (NMR) | 0.86 | a,b |
| Protonated toluene/toluene | $HPhCH_3^+$ | $PhCH_3$ | ???? | | c |
| **Acidic Electrolytes** | and | weak bases | | | |
| protonated pentafluoropyridine | $C_5F_5NH^+$ | $C_5F_5N$ | -13 | 0.77 | d |
| Sulfuric acid | $H_2SO_4$ | $HSO_4^-$ | -12 | 0.71 | a,b,e |
| Nitric acid | $HNO_3$ | $NO_3^-$ | -1.3,-1.5 | 0.08 | f, g |
| protonated 2-fluoropyridine | $C_5H_4FNH^+$ | $C_5H_4FN$ | -0.43 | 0.025 | g |
| Trifluoroacetic acid, TFA | $CF_3COOH$ | $CF_3COO^-$ | -0.25 | 0.015 | |
| **Neutral Electrolytes:** | | | | | |
| **Hydronium/Water** | $H_3O^+$ | $H_2O$ | -1.7, 0.00 | 0.00 | |
| Diphenylammonium/DPA | $Ph_2NH_2^+$ | $Ph_2NH$ | 0.79 | 0.047 | |
| Hydrogen sulfate | $HSO_4^-$ | $SO_4^{2-}$ | 1.92 | | |
| Formic acid | HCOOH | $HCOO^-$ | 3.75 | -0.19 | |
| Phenylamine (as base) | $C_6H_5NH_3^+$ | $C_6H_5NH_2$ | 4.6 | -0.22 | |
| Acetic acid | $CH_3COOH$ | $CH_3COO^-$ | 4.75 | -0.27 | |
| N,n-diformylaminium/ine | $H_2N(CH{=}O)_2^+$ | $HN(CH{=}O)_2$ | ?? | | |
| **Hydrogen sulfide** | $H_2S$ | $HS^-$ | 7 | -0.41 | |
| Bis- (trimethylsilyl)-ammonium | $((CH_3)_3Si)_2NH_2^+$ | $((CH_3)_3Si)_2NH$ | 7.5 | | |
| Ammonium/Ammonia | $NH_4^+$ | $NH_3$ | 9.23 | -0.55 | |
| Boric acid | $H_3BO_3$ | $H_2BO_3^-$ | 9.24 | | |
| Ethylammonium/Ethylamine | $C_2H_5NH_3^+$ | $C_2H_5NH_2$ | 10.63 | | |
| Dicyanomethane (as acid) | $H_2C(CN)_2$ | $HC(CN)_2^-$ | 11.0 | -0.63 | |
| di-isopropylammmonium/DPA | $[H_2N(C(CH_3)_3)_2]^+$ | $HN(C(CH_3)_3)_2$ | 11.1 | | |
| Bisulfide/Sulfide | $HS^-$ | $S^{2-}$ | 12.4 | | |
| Dihydrogen borate anion | $H_2BO_3^-$ | $HBO_3^{2-}$ | 12.9 | | |
| Guanidinium/Guanidine | $(NH_2)_3C^+$ | $(NH_2)_2C{=}NH$ | 13. | | |
| Water/ hydroxyl | H2O | OH- | 14 | -0.83 | |

**References: a,** Olah, G. A.; et al., Superacid Chemistry, 2nd ed.; Wiley and Sons, 2009, **b -g , below next table**

| **Basic Electrolytes:** | **and** | **Strong bases** | | |
|---|---|---|---|---|
| | occupied | vacant | ΔpKa* | ε/eV * |
| Water (Standard state) | $H_2O$ | $OH^-$ (Na+, Li+)* | 14 | -0.83 |
| Water/Hydroxyl | $H_2O$ | OH- | 15.7 (Evans (f) | |
| DBU⁺/DBU* | $C_9H_{16}N_2H^+$ | $C_9H_{16}N_2$ | 14 (f,g) 13.5 (h) | |
| N,N,diformylamine/ -ide | $HN(CH{=}O)_2$ | $^-N(CH{=}O)_2$ (Na+)* | 14,15 | -0.83 |
| 1,5,7triazobcyclo[440] | TBDH | TBD | 16 | |
| Phosphazene* | $P_2$-EtH⁺ | $P_2$-Et | 15.2,() 20 ±2(h) | -0.94, -1.18 (h) |
| Acetylene (as acid) | HCCH | $HCC^-$ ($Na^+$)* | 20 (Evans,(f)) 25 (g) | -1.18,-1.66 |

6 5

| | | | | |
|---|---|---|---|---|
| phenylacetylene/acetylide<br>Diphenylamine (as acid)<br>Acetylide/carbide<br>Diformylamide as acid | PhCCH<br>$HNPh_2$<br>$HCC^-$ 4<br>$^-N(CH{=}O)_2$ (Na+) | $PhCC^-$ $(Na^+)$*<br>$Ph_2N$-<br>$C_2^{2-}$ $(Na^+)$*<br>$^{2-}N(CH{=}O)(C{=}O)$ (Na+) | 23 (Evans)<br>25 (Evans,DMSO)<br>20.9 (ΔG)<br>??? | -1.36<br>-1.48<br>-1.24 (ΔG) |
| **Hydroxyl/Oxide** | $OH^-$ (Na+)<br><br>$OH^-$ $(Li^+)$ | $O^{2-}$ $(Na^+)$<br>5<br>$O^{2-}$ $(Li^+)$ | **25 (28)**<br>27((ΔG, $Na^+$)<br>20.9(ΔG) | -1.66, ref.<br>-1.58 (ΔGº)<br>-1.24(ΔG) |
| bis(triMesilyl)amine/amide<br>Formylamine/formylamide<br>dialkylamine/dialkylamide | $((CH_3)_3Si)_2NH$<br>$(CHO)NH_2$<br>$HN(isopropyl)_2$ | $((CH_3)_3Si)_2N^-(Li^+)$*<br>$(CHO)NH^-$<br>$N(isopropyl)_2^-$ $(Li^+)$* | 26<br><br>36 (THF) , 40 | 1.534 |
| **Ammonia /Amide**<br><br>**Phenylamide as acid** | $NH_3$<br><br>$PhNH^-$ | $NH_2^-$ $(Na^+)$*<br>$NH_2^-$ $(Li^+)$<br>$PhN^{2-}$ | 30.6(ΔG)<br>~25 (ΔG) 38 (Evans)<br>unknown | -1.81 (ΔG)<br>-1.64 (ΔG) |
| Amide/imide. | $NH_2^-$ | $NH^{2-}(Na^+)$* | 34 ? | ? |
| **Hydrogen/hydride** | $H_2$ | $H^-$ (Na+)<br>$H^-$ (Li+)<br>4 | 33.1 (ΔG)<br>37.5 (ΔG) (~36 Evans(f) | -1.96 (ΔG)<br>-2.22 (ΔG) |
| **Imide/nitride** | $NH^{2-}$ | $N^{3-}(Li^+)$* | 40.37((ΔG)<br>42.4(ΔGº $H_2O$ | -2.39 (ΔG)<br>-2.5 (ΔG $H_2O$) |
| **toluene** | $PhCH_3$ | $PhCH_2^-$ | 41 (ref g) | |
| Methane/methide<br>(methyllithium) | $CH_4$ | $CH_3^-$ $(Li^+)$* | ~48<br>(48 Evans) | No $ΔG_f$º<br>available |

ionic liquids yet produced, (the case of transfer of a proton from acetic acid to DBU(*6*)), which should lie midway between the levels HOAc/OAc- (4.7) and DBU+/DBU (see below), would only be ~9 on the water scale. So, if basicity is taken to mean having an effective $pK_a$ value for the PIL higher than 14, then the latter PIL which achieved high ionicity(*6*), did not achieve basicity.

The basis for this statement is best understood by considering the formation of the protic ionic liquid as a product of a titration of the neat Brønsted acid by the neat Brønsted base as performed by Kanzaki et al. (18), and then considering (using examples illustrated by arrows in Figure 1), the other proton transfer processes that are possible. We use Figure 1 to illustrate three of these alternative processes, and Scheme 1 to complete the list of the six possible cases. The important thing to realize is that by choosing as proton acceptor, an anionic moiety rather than a molecule, an ionic liquid product of much lower proton activity than otherwise accessible can, in principle at least, be generated. The ionic liquids we discuss will in most cases not have proton donor capability, but will rather act as a proton acceptors, i.e.  reducing agents to any chemical of higher proton activity with which  they might find themselves in contact (kinetics allowing).

There is now the problem of assessing what are the energies, and $pK_a$ values on a scale consistent with those in Chart 1. In the next section we approach this challenge in a simple but unambiguous manner which we have not seen used elsewhere. The approach is suitable for solvent-free protic systems and ties the proton activities directly to the water scale. While it involves some obvious approximations, the level of approximation can be assessed and is not large.

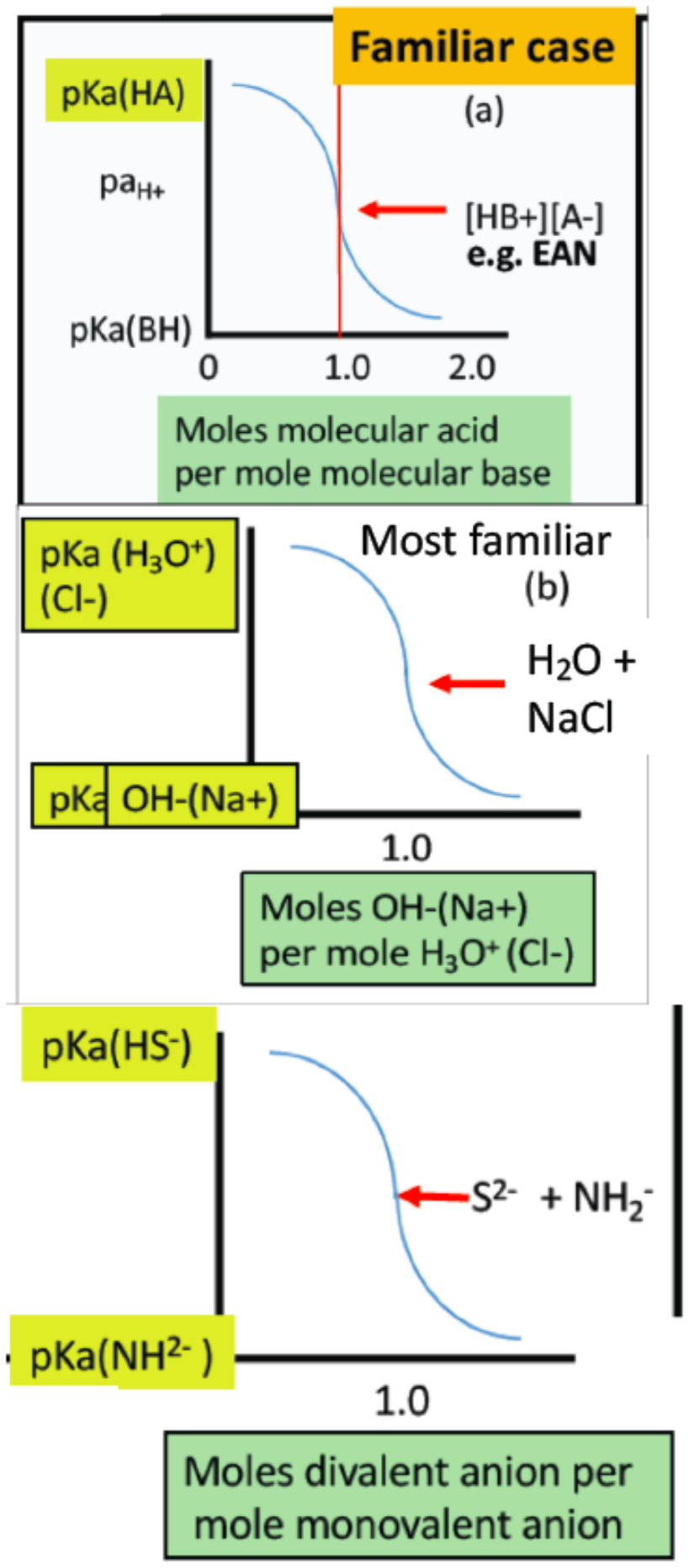

**Scheme 1: Six possible proton transfer processes**

(a) Molecular acid to molec. base e.g. $HNO_3$ + EA (arrows 1)
(b) protic cation to anionic base e.g. $H_3O^+ + OH^-$ (and arrow 5)
(c) monovalent anion to divalent anion e.g. $NH^{2-} + S^{2-}$ (arrow6)
(d) Protic cation to molec. base e.g. $Ph_2NH_2^+$ + Gdn (arrow2)
(e) molecular acid to anionic base e.g. $HNPh_2 + N^{3-}$ (arrow 4)
(f) protic anion to molecular base e.g. HSO4- + Gdn (arrow 3)

**Figure 1.** Three cases of acid-base titrations, the first two being familiar, to illustrate the pKa value of the product of the titration relative to those of the starting couples. The third case illustrates how a 1:1 mixture of anions of the appropriate character can incorporate anions of high basicity at the equimolar composition. The stable pair of anions indicated would need charge compensation by cations of non-reducible character.

2.

3. **Assessing the approximate proton potential of some important donor-acceptor couples from available thermodynamic data.**

As a first example of how to estimate the energy, and hence $pK_a$ value, of a key occ/vac level, referenced to $H_3O^+/H_2O$ as zero, we take the case of proton transfer between water and oxide anion) levels (see blue arrow on Chart 2). Water is now the acid i.e. proton donor

Written as a chemical equation, this would be

$H_2O + O^{2-} = 2OH^-$ (1)

and considering that the anions do not exist without charge compensating partners, this can be expanded to read

$H_2O + Na_2O = 2NaOH$ (2)

Since the molar free energies of formation of all components of Eq. (2) are known(*25*) it is a simple matter, using the free energy equivalent of Hess' Law, to find the free energy change in the process. We locate the molar free energies of the three compounds under the components so we can add them in the manner dictated by Hess's law

$H_2O + Na_2O = 2NaOH$ (2)

+237.1 + 416 2x-379.4 all in kJ/mole:

The addition yields $\Delta G^o$ = -146.8 kJ for the two moles of protons transferred or -73.4 kJ/mole of protons transferred.

This can be converted to an energy in electron volts, ε eV per proton transferred (as was preferred by Gurney), using the conversion formula 100kJ/mole ≡ 1.036 eV per elementary charge* i.e. per proton, thus ε = 0.76 eV. (*Conversion factors are 6.24 x$10^{18}$ $eVJ^{-1}$ and 6.028 x $10^{23}$electrons (mole of electrons)$^{-1}$)

Alternatively, one can use the macroscopic formula

$E^o = -\Delta G^o/F$ (where F is the Faraday) to obtain the potential difference in volts between the $H_2O/OH^-$ and the $OH^-/O^{2-}$ levels as 73,400 $Jmole^{-1}$/ 96,500 $Cmole^{-1}$ = 0.76 J/C = 0.76V.

.

To obtain the value for the column 4 potential in Table 2, the value for the potential difference between $H_3O^+/H_2O$ and $H_2O/OH$ levels, viz., -0.83 V, must be added, yielding -1.59 V for the potential of the $OH^-/O^{2-}$ level. We convert this to a $pK_a$ value for the level, using the relation between free energy gap $\Delta\varepsilon^o$ and $\Delta pK_a$, which is

$\Delta\varepsilon^o = -2.303(RT/nF)\Delta pK_a$ (3)

So using the familiar value of 2.303RT/F (= 0.059 for a single charge transfer at 25ºC), we obtain $\Delta pK_a = -\Delta\varepsilon^o/0.059$. This yields $\Delta pK_a$ = 27, i.e. a $pK_a$ value of 27 for the $OH^-/O^{2-}$ level. This is deep in the basic domain.

Before considering this and other cases, and their related anions for high basicity ionic liquids, let us reassure ourselves that the approach leading to this $pK_a$ assignment is sound. We can accomplish this objective by using the standard free energies of formation of $H_2SO_4$, $Na_2SO_4$, NaOH and $H_2O$ to obtain the potential and $pK_a$ value for the better-known, indeed seminal, level $H_2SO_4/HSO_4^-$ in the strong acid domain, also keeping a water reference. This is done by assessing $\Delta G^o$ for the neutralization of $H_2SO_4$ by NaOH, using Eq. (4) which is descriptive of two protons falling from the $H_2SO_4/HSO_4^-$ level down to the $H_2O/OH^-$ level in Chart 1.

$H_2SO_4$ + 2NaOH = $2H_2O$ + $Na_2SO_4$ (4)

+ 689.9 + 2 x 379.4 - 2 x 237.1 -1270.2 (ΔG = -295. 7) Na

Other cases to be considered will be those where Na cations are replaced by Li+ or K+

| | | | |
|---|---|---|---|
| 439 | -1321.7 | (-286.8) | Li |
| 378.7 | -1321.4 | (-347.6) | K |

Here we have again used the pure substance values, not the aqueous solution values, for the free energies of formation since we are going to be comparing with the Hammett acidity function for the pure acid $H_2SO_4$. These lead to $\Delta G^o$ = -295.7 kJ/mole for the two proton process, or 147.9 kJ/mole per proton, and $\varepsilon^o$ = 1.53eV above the $H_2O/OH^-$ level. Since the latter is 0.83V below the $H_3O^+/H_2O$ reference level we need to subtract this amount from the 1.53V total to obtain the energy of the $H_2SO_4/HSO_4^-$ level which is thus predicted to be 0.70 eV. This is to be compared with the value 0.71V that we obtained from the experimentally determined H(0) value for pure $H_2SO_4$ (21) see Chart 1. The corresponding $pK_a$ value for this level is -0.70/0.059, or -11.9, to be compared with -12.0 from H(0) in Chart 1.

The agreement is too good to be true, and we should recognize that the numbers obtained will be different if we should use the free energies of formation of LiOH in place of NaOH, and $Li_2SO_4$ in place of $Na_2SO_4$, in the calculation. Fortunately, values for both are available and $\Delta G$ is almost unchanged at 286.8 kJ/mole, implying $\varepsilon^o$ = 1.48V, or 0.65V after subtraction of the same 0.83V. For this proton potential we obtain a $pK_a$ value of -11.0 for the $H_2SO_4/HSO_4^-$level. However, free energy of formation data are also available for KOH and $K_2SO_4$ , and in this case the free energy of the reaction proves to be somewhat larger, at 347.6 kJ/mole, implying $\varepsilon^o$ = 1.79V and eventually $pK_a$ for the $H_2SO_4/HSO_4$ level, rather larger at -16.3. Such differences are not too surprising insofar as the free energies of two of the four reaction participants have their own special crystal structures, hence lattice energies, that play a role in determining their free energies of formation. Indeed, even in the liquid state, the cation that charge-neutralizes a given anion must be expected to play a significant role in determining the chemical properties, which is not the case when the ions are dispersed in a solvent, e.g. 55.5 molecules of water per ion, as in a one molar aqueous solution. Nevertheless, there can be little doubt that the simple thermodynamic route is putting us in the right zone of proton potentials for the ionic liquids of interest to us.

With this in mind we return to the basic range to obtain some other landmark proton potential levels in the basic range using thermodynamic data for the relevant components that are available in the literature. Again there are cases for which data for both Li and Na salts are available, and substantial (and unpredictable) differences are found depending on which alkali cation is involved.

Two other levels deeper in the basic range that we can assess are the ammonia/amide ($NH_3/NH_2^-$) level, (the more basic analog of the $H_2O/OH^-$ level), and the hydrogen/hydride level ($H_2/H^-$), both involving gaseous proton donors. For the first we invoke the water hydrolysis reaction

$$H_2O(l) + LiNH_2(s) \rightarrow NH_3(g) + LiOH(s) \qquad (5)$$

Using $\Delta G^o_f$ = -140.6 kJ mol$^{-1}$ for $LiNH_2$(s), -16.5 kJ mol$^{-1}$ for $NH_3$(g), and -439 kJ mol$^{-1}$ for LiOH(s), we obtain a value of ΔG for the single proton transfer process of -77.9 kJ mol$^{-1}$, or Δε = 0.81. This corresponds to a proton potential for the $NH_3/NH_2^-$ (Li+) couple that lies 0.81 eV below the $H_2O$/OH-level hence 1.63 eV below the $H_3O^+/H_2O$ reference level. The corresponding $pK_a$ value is 25, similar to that for the hydroxyl/oxide level. However if the $\Delta G^o_f$ values for $NaNH_2$(s) and NaOH(s) are used in place of those for $LiNH_2$ and LiOH in Eq. (5), the values for Δε and $pK_a$ change to the more basic numbers, -1.81V and 30.6. Both sets of numbers are entered in Chart 2.

For the second case, free energy of formation data for both LiH and NaH are also available (-67.6 kJ/mole for LiH, and -33.5kJ/mole for NaH), and the reactions with water, forming hydroxides and releasing $H_2$(g) can be assessed and lead to the even more basic level numbers listed in Table 2, where the derived $pK_a$ values of 33.1 and 37.5 are listed.

Finally, the free energy of formation of $Li_3N$ is reported to be -138 kJ mol$^{-1}$ and an equation for its reaction with water yielding nitrogen and LiOH

$$3H_2O(l) + Li_3N(s) \rightarrow NH_3(g) + 3LiOH(s) \qquad (6)$$

is found to have a free energy change of -484 kJ mol$^{-1}$ for 3 moles of protons transferred or 161.3 kJ mol$^{-1}$. This implies an energy level for the first of the steps in the series leading to $NH_3$ and NaOH, namely $N^{3-}/NH^{2-}$, that lies 2.5 eV below the reference $H_3O^+/H_2O$ level, with a corresponding pKa value of 40.4, as entered in the Table 2.

It is worth noting that not even the reaction of metallic sodium with water has a larger free energy change per mole of protons transferred (to form hydrogen and NaOH),

$$Na(s) + H_2O(l) \rightarrow \tfrac{1}{2} H_2(g) + NaOH(s) \qquad \Delta G = \text{only } -142 \text{ kJ/mole.} \quad (7)$$

## 3. Making ionic liquids with $pK_a$ values in the new high basicity ranges.

It is one thing to identify the existence of an extended basicity range, and a quite different thing to populate it with actual examples of ionic liquids with anions of very high basicity. We leave this for separate papers(*26, 27*), but mention here that anyone can demonstrate the existence of a new and basic member of the anion family of general formula $NX_2^-$. The $pK_a$ range of this remarkable family extends from -18 in the case of the $NTf_2^-$ anion (of the super acid $HNTf_2$ also known as HTFSI), through 0 to about +40 for the case of the amide ion itself, $NH_2^-$, (as characterized in the previous section.) The new member is the N,N,diformylamide anion, $N(CHO)_2^-$, which is produced as the anionic component of a mobile ionic liquid by centrifuging off the NaCl product of a metathetical solid-solid reaction that occurs on mixing sodium N,N,diformylamide and pyrrolidinium chloride. Provided temperature is kept below 60ºC, the liquid product, designated $[P14]^+[N(CHO)_2]^-$, can be identified by NMR as containing only pyrrollidinium cations and N,N diformylamide anions. It has conductivity and viscosity comparable with those of P14dicyanamide (*28*), and a $pK_a$ value of about 14 (much more basic than DBU acetate). The ionic liquid made in a similar manner but with the HMDS anion, $\{(CH_3)_3Si\}_2N^-$, must have a $pK_a$ value of about 26. Details of these and other preparations will be given in separate papers(*26,27*). It will be exciting to check the basicities of these new liquids using the redox potential of the Fe(II)/Fe(III) couple that showed such a systematic variation of emf vs Ag/AgCl reference in a previous study of oxidic solvent anion basicity effects(10).

## Acknowledgements

This work has been supported by the DOD Army Research Office under Grant No. W911-NF19-10152